\documentclass[final,5p,times,twocolumn]{elsarticle}

\usepackage{amssymb}
\usepackage{amsmath}
\usepackage{url}
\usepackage{amsfonts}
\usepackage{amssymb}
\usepackage{multirow}
\usepackage{verbatim}
\usepackage{color}
\usepackage{algorithm}
\usepackage[noend]{algpseudocode}
\usepackage{appendix}
\usepackage{graphicx}
\usepackage{lipsum}

\begin{document}

\begin{frontmatter}



\title{Parallel Algorithms for Constructing Data Structures for Fast Multipole Methods}
\author[CS,UMIACS]{Qi Hu \corref{huqi}\fnref{hq}}
\author[UMIACS,Fantalgo]{Nail A. Gumerov}
\author[CS,UMIACS,Fantalgo]{Ramani Duraiswami}
\address[CS]{Department of Computer Science, University of Maryland, College Park}
\address[UMIACS]{University of Maryland Institute for Advanced Computer Studies (UMIACS)}
\address[Fantalgo]{Fantalgo LLC, Elkridge, MD}
\fntext[hq]{please contact the corresponding author via
  email: \tt{huqi@cs.umd.edu} or phone:+13014051207 (fax:+13013149658)}
\begin{abstract}
We present efficient algorithms to build data
structures and the lists needed for fast multipole
methods. The algorithms are capable of being efficiently implemented on both serial, data parallel GPU and on distributed architectures. With these algorithms it is possible to map the  FMM efficiently  on to the GPU or distributed heterogeneous CPU-GPU systems. Further, in dynamic problems, as the distribution of the particles change, the reduced cost of building the
data structures improves performance. Using these algorithms, we demonstrate example high fidelity simulations with large problem sizes by
using FMM on both single and multiple heterogeneous computing facilities equipped with multi-core CPU and many-core GPUs.
\end{abstract}


\end{frontmatter}

\section{Introduction}
\label{sec1}

$N$-body problems evaluate the weighted sum of a kernel
function $\Phi(\mathbf{y},\mathbf{x}_i)$ centered at $N$ source locations
$\{\mathbf{x}_{i}\}$ for all $M$ receiver locations
$\{\mathbf{y}_{j}\}$ with the strengths $q_{i}$
(Eq.~\ref{fmm_eq1}). They can also
be viewed as dense $M\times N$ matrix vector products. Direct
evaluation of this method on CPU has the quadratic $O(NM)$
complexity. Such direct evaluations cannot be scaled to large sizes
required by high fidelity simulations.   
\begin{equation}
\phi (\mathbf{y}_{j})=\displaystyle\sum_{i=1}^{N}q_{i}\Phi (\mathbf{y}_{j}-
\mathbf{x}_{i}),\ \ j=1,2,\ldots ,M,\ \ \mathbf{x}_{i},\mathbf{y}_{j}\in
\mathbb{R}^{d},
\label{fmm_eq1}
\end{equation}
Hardware acceleration, such as \cite{Nyland07:GPUGEM3} using the GPU and  \cite{Gualandris2007:PAD} using
specially constructed hardware called the ``Gravity Pipe''(GRAPE), can only speedup the sum to
some extent but do not address its quadratic complexity. 

An alternative way to evaluate such sums for particular kernels is to use
\textit{fast approximation algorithms}, for example, the Fast Multipole
Method \cite{Greengard87:JCP}, the Barnes-Hut Method
\cite{Barnes86:Nature} and the Particle-Mesh Methods
\cite{Darden93:JCP}, which have lower asymptotic complexity when they are applicable. Since
the FMM can achieve linear complexity but achieve
guaranteed accuracy up to  machine precision, we
only focus on this method in this paper, however the data structure
techniques used here may find application in the other fast algorithms
also, and indeed wherever computations involve particles. 

The FMM exactly computes near-field interactions but approximates far-field
interactions to a specified tolerance $\epsilon $. It splits the
summation in Eq.~\ref{fmm_eq1} into near and far fields as
\begin{equation}
\phi (\mathbf{y}_{j})=\displaystyle\sum_{\mathbf{x}_{i}\in \Omega
  (\mathbf{y}_{j})}q_{i}\Phi
(\mathbf{y}_{j}-\mathbf{x}_{i})+\sum_{\mathbf{x}_{i}\not\in \Omega
  (\mathbf{y}_{j})}q_{i}\Phi (\mathbf{y}_{j}-\mathbf{x}_{i})
\label{fmm_eq2}
\end{equation}
\noindent for $j=1,2,\ldots,M$ where $\Omega $ is the neighborhood
domain. The first term on the right hand side of Eq.~\ref{fmm_eq2} can be computed exactly at $O(N)$
cost given a fixed cluster size, i.e., the maximal number of
data points inside any neighborhood domain. To approximate the second term, the kernel function is factored
into an infinite sum, which is truncated at $p$ terms according to the
required accuracy, by using singular (multipole) spherical basis functions, $S_{l}$,
and regular (local) spherical basis functions $R_{l}$. These
factorizations can be used to separate the kernel computations involving
the points sets either $\{\mathbf{x}_{i}\}$ or $\{\mathbf{y}_{j}\}$, and
consolidate operations for many points as
\begin{equation}
\begin{array}{l}
\displaystyle\sum_{\mathbf{x}_{i}\not\in \Omega
(\mathbf{y}_{j})}q_{i}\Phi (\mathbf{y}_{j}-\mathbf{x}_{i})=\displaystyle\sum_{\mathbf{x}_{i}\not\in \Omega
(\mathbf{y}_{j})}q_{i}\sum_{l=0}^{p-1}S_{l}(\mathbf{y}_{j}-\mathbf{x}
_{\ast })R_{l}(\mathbf{x}_{i}-\mathbf{x}_{\ast }) \\
=\displaystyle\sum_{l=0}^{p-1}S_{l}(\mathbf{y}_{j}-\mathbf{x}_{\ast
})\sum_{\mathbf{x}_{i}\not\in \Omega
(\mathbf{y}_{j})}q_{i}R_{l}(\mathbf{x}
_{i}-\mathbf{x}_{\ast })=\displaystyle\sum_{l=0}^{p-1}C_{l}S_{l}(\mathbf{y}_{j}-\mathbf{x}_{\ast
}),
\end{array}\label{fmm_eq3}
\end{equation}
\noindent The $p$ coefficients $C_{l}$ for all $\mathbf{x}_{i}$ are
built in $pN$ operations and then they can be used in the evaluation
at all $\mathbf{y}_{j}$ in $pM$ operations. This
approach reduces the cost of evaluating the far-field contributions as well as the
memory requirement to $O(N+M)$.

Because the factorization in Eq.~\ref{fmm_eq3} is not global, the
split between the near- and far-fields must be managed, which requires
appropriate data structures and the use of a variety of representations for the function. The
efficiency with which the data structures are constructed is very important for dynamic problems
since the source and receiver points change their positions at every
time step. We propose novel parallel algorithms for the data
structures for both single and multiple heterogeneous nodes.

\subsection{Well-separated Pair Decomposition}
The need to construct spatial data structures arise from a need to provide an  error
controlled translation of the FMM function representations (discussed below in Section \ref{Intro:MLFMMDS}). This is achieved using  a \textit{well-separated pair decomposition}
(WSPD), which is itself useful for solving a
number of other geometric problems \cite[chapter 2]{Samet2005:FMMDS}. In the
context of FMM, given the distance between the two sphere centers $d$,
with radii are $r_A$ and $r_B$ respectively, the translation error 
$\varepsilon$ (from $c_A$ to $c_B$) is bounded by
\begin{equation}
\varepsilon(p) < C\eta^p{(r_A,r_B)}, \quad \eta(r_A,r_B)=\frac{\max(r_A,r_B)}{d-\min(r_A,r_B)} \label{wsperr}
\end{equation}where $p$ is the truncation number. Note that the $p$ is
determined based on the worst case. In the FMM data structures,
this WSPD is realized by the octree spatial decomposition (Refer to
theoretical results from \cite{Anderson1999:TDS}).

\subsection{The Baseline FMM Algorithm}
\label{sec2}

The FMM was first introduced by Greengard
and Rokhlin in \cite{Greengard87:JCP} and has been identified as one of
the ten most important algorithmic contributions of the 20th century
\cite {Dongarra00:CSE}. 


The multi-level FMM (MLFMM) puts sources into hierarchical space boxes
and translates the consolidated interactions of sources into
receivers. For the convenience of presentation, we
call a box containing at least one source
point a \textit{source box} and a box containing at least one receiver
point a \textit{receiver box}. The FMM algorithm can be summarized as four main parts: the initial
expansion, the upward pass, the downward pass and the final
summation.

\begin{enumerate}
\item  \textbf{Initial expansion (P2M)}:

\begin{enumerate}
\item  In the finest level $l_{max}$, all the source data points are
  expanded at their box centers to obtain the far-field $\mathcal{M}$
  expansion coefficients $\{\mathcal{C}_n^m\}$ over $p^{2}$ spherical basis
  functions. 
\item  The obtained $\mathcal{M}$-expansion from all source points
  within the same boxes are consolidated into a single expansion at
  each box center.
\end{enumerate}

\item  \textbf{Upward pass (M2M)}: For levels from $l_{max}$ to 2, the
  $\mathcal{M}$ expansion coefficients for each box are evaluated via the
  multipole-to-multipole ($\mathcal{M}|\mathcal{M}$) translations from
  the source boxes to their parent source box. All these translations
  are performed in a hierarchical order from bottom to top via the octree.

\item  \textbf{Downward pass}: For levels from 2 to $l_{max}$,
  each receiver box also generates its local or $\mathcal{L}$
  expansion in a hierarchical order from top to bottom via the octree.

\begin{enumerate}
\item  M2L: Translate multipole $\mathcal{M}$ expansion coefficients
  from the source boxes of the same level belonging to the receiver
  box's parent neighborhood but not the neighborhood of that receiver
  itself, to local $\mathcal{L}$ expansion via multipole-to-local
  ($\mathcal{M}|\mathcal{L}$) translations then consolidate the
  expansion coefficients.

\item  L2L: Translate the $\mathcal{L}$ expansion coefficients (if the
  level is 2, then these expansions are set to be 0) from the parent
  receiver box center to its child box centers and consolidate with
  the same level multipole-to-local translated expansions.
\end{enumerate}

\item  \textbf{Final summation (L2P}: Evaluate the $\mathcal{L}$
  expansion coefficients for all the receiver points at the finest
  level $l_{max}$ and performs a local direct sum of nearby source
  points within their neighborhood domains.
\end{enumerate}

Note that the local direct sum is independent of
the far-field expansions and translations, thus may be scheduled
on different computing hardware concurrently for high performance
efficiency. Moreover, it is important to balance costs between these pairwise kernel sums and
the hierarchical translations to achieve high computation throughput
and proper scaling. Besides those algorithmic considerations, there is
another vital factor to achieve such desired high efficiency: low data
addressing latency. In our implementation, both translations and local
direct sums have their special auxiliary {\it interaction lists} used
to address data directly. Therefore,
the FMM algorithm requires the following special data structures:\begin{enumerate}
\item octree to ensure WSPD that ensures error bounds.
\item interaction lists for fast data addressing.
\item the communication management structures.
\end{enumerate}The construction of these data structures must be done
via algorithms that have the same overall complexity with the
summation. 

\subsection{Treecode and Its Data Structures}
Similar to the FMM, there is also another well-known fast $N$-body
simulation algorithm, {\em Barnes-Hut-Method} \cite{Barnes86:Nature},
which uses the similar spatial data structures as FMM and is often
called a {\em treecode}. As in the FMM, the whole space is hierarchically
subdivided via an octree. Each spatial box has an pseudo-particle that
contains the total mass in the box located at the center of mass of
all the particles it contains. Whenever force on a particle is
required, the tree is traversed from the root. If a certain box is far
away from that particle, the pseudo-particle is used to approximate
the force induced by that box, otherwise it is subdivided again or is
processed particle--by--particle directly. The complexity of treecode is
in $O(N)$. However, unlike FMM, the control on the accuracy is less precise.

In the most recent GPU treecode development \cite{Bedorf2012:JCP},
algorithms for the octree traversing, particle sorting and data compaction (skip empty boxes) on
GPU based on the {\it cuda scan} algorithm \cite{CudaScan}. Such
algorithms, are similar to the approaches in Section~\ref{SINGLEFMMDS:Sort}
and which we first presented in \cite{Qi2011:AHS}. The other work
in the treecode space that is similar, is \cite{Ajmera2008:FPG}, in which a GPU-based construction of space
filling curves (SFC) and octrees were presented.  

\subsection{Multi-Level FMM Data Structures}
\label{Intro:MLFMMDS}

\begin{figure}[!t]
\begin{center}
\includegraphics[width=0.4\textwidth]{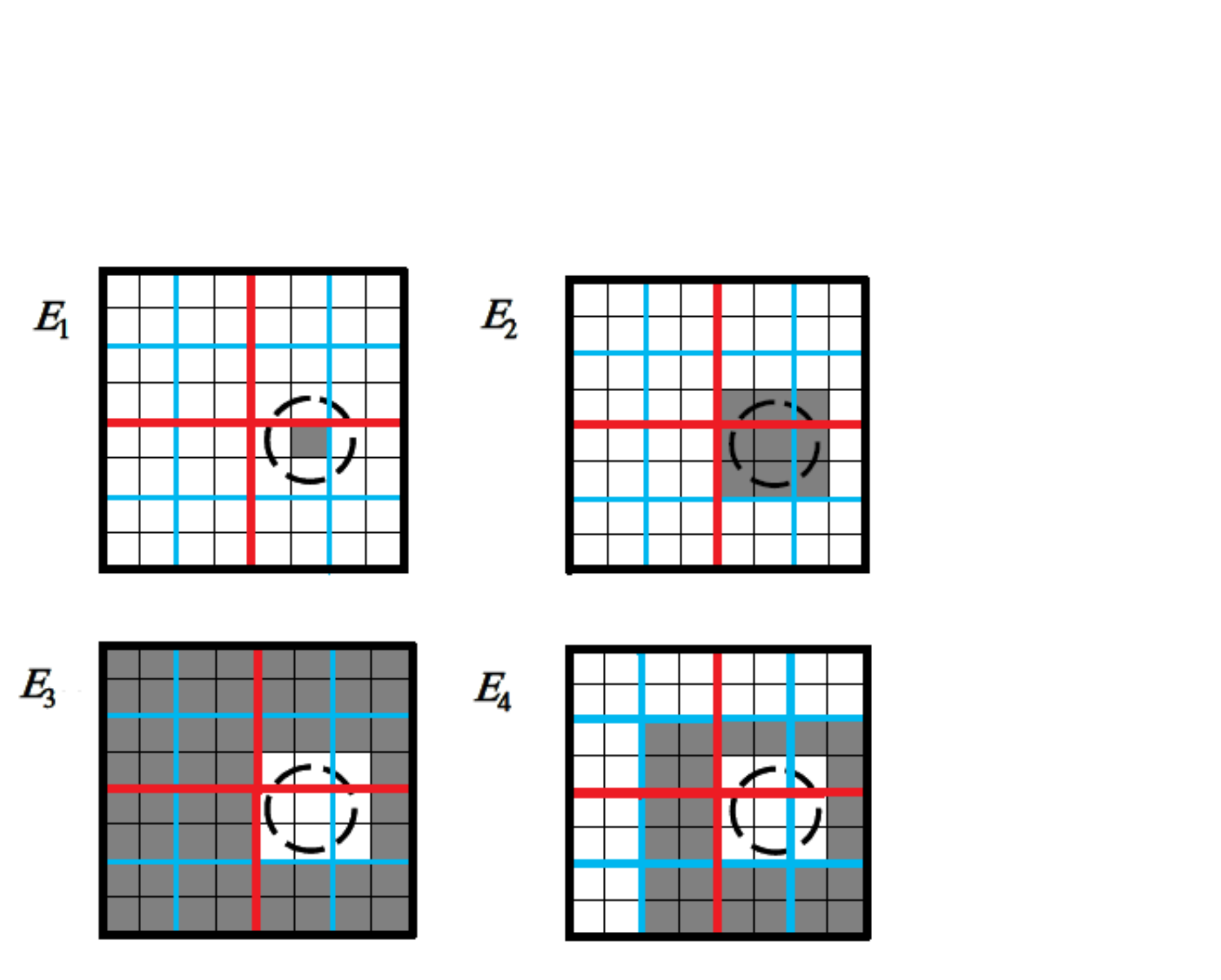}
\end{center}
\caption{$E_{1},E_{2},E_{3},E_{4}$ neighborhoods of dimension 2: red
division at level 1; blue division at level 2; black division at
level 3. The shaded box in $E_{1}$ sub-figure has its Morton index as
40 at level 2.  } \label{fmmDS}
\end{figure}

Assume all of the data points are already scaled into a unit cube. The WSPD is
recursively performed by subdividing the cube into subcubes (spatial
boxes) via an octree until the maximal level, $l_{max}$, or the tree
depth, is achieved (The level $l_{max}$ is chosen such that the
computational costs of the local direct sums and the far-field
translations can be balanced to the extent possible). To guarantee separation of spatial data points by these subcubes
and their minimal bounding spheres, we need to introduce several different space neighborhood
domains \cite{Gumerov2004:FMMBook}. Given
each spatial cubic box with the Morton index
\cite{Samet2005:FMMDS,Morton66:GSSS} $n=0,\ldots,2^{ld}$ at level
$l=0,\ldots ,l_{max}$ in $d$ dimensions,  

\begin{enumerate}
\item  $E_1(n,l) \subset \mathbb{R}^d$ denotes the spatial points
  inside the box $n$ at level $l$. We call these boxes as source
  or receiver box with index $n$ at level $l$. 
\item  $E_2(n,l) \subset \mathbb{R}^d$ denotes the spatial points
  in the neighborhood of the box with index $n$ at level $l$
  (``neighborhood'' means all its immediate neighbor boxes). This
  list is used for local direct summations for $E_1(n,l)$.  
\item  $E_3(n,l) = \overline{E_2(n,l)}\subset \mathbb{R}^d$ denotes
  spatial points outside the neighborhood of the box $n$ at level
  $l$. This is the complement of $E_2(n,l)$.
\item  $E_4(n,l) = E_2($\texttt{ParentIndex}$(n),l-1)\backslash
  E_2(n,l) \subset \mathbb{R}^d $ denotes spatial points inside the
  neighborhood of the parent box \texttt{ParentIndex}($n$) at level
  $l-1$ but which do not belong to the neighborhood of box $n$ at level
  $l$. These are {\it interaction boxes} whose contributions are
  accounted for by M2L translations for $E_1(n,l)$. 
\end{enumerate}

Consider any box $B$ with Morton index $n$ at level $l$ (see Fig.~\ref{fmmDS}). All
the translation operations are performed box by box so the source data
have to be viewed as spatial boxes but not individual points. All the receiver data
points inside $B$ can not be well
separated with all the source boxes inside $E_2(n,l)$. Hence
$E_2(n,l)$ is used to compute the near-field sum. Due to hierarchical
translations, all the source boxes outside $E_4(n,l)$ have already been
translated to $B$'s center at the previous level. Thus only the
influence of the remaining source data needs to be translated. These
are located in its $E_4(n,l)$ domain, which corresponds to the most time consuming $M|L$ translation
to $B$.


In \cite{Gumerov03:DS} and \cite{Gumerov2004:FMMBook}, they described
those FMM related octree data structures and their implementations in
details. Similar work on such hierarchical spatial data structures can be
found \cite{Sevilgen1999:UDS} and \cite{Hariharan2005:EPA}. 


In the literature, the data structure research mainly focuses on load balancing and data partition. In
\cite{Greengard1990:CMA}, several opportunities
for parallelism in the FMM were discussed and it was shown that it is possible to apply FMM on
both shared memory or distributed architectures. Compared to  later
work, the data distribution method in this pioneering paper was simple,
perhaps not practical in many applications. In \cite{Singh1993:SC93},
an efficient parallel adaptive FMM with a ``costzones'' partition
technique was developed based on data locality. A multi-threaded tree
construction was implemented in \cite{Mahawar2004:HiPC04}. However, in these papers the data structures were not built in parallel,  i.e, the local tree of each node was
built by a single processor.  In \cite{Ying2003:SC03} and \cite{Yokota09:CPC} they separated the
computation and communication to avoid synchronization during the
evaluation passes. Ref.~\cite{Lashuk2009:SC09} extended the work
of \cite{Ying2003:SC03} by providing a new parallel tree construction and a novel communication scheme, which scaled up to
billion size problem on 65K cores. But all the GPUs were only used for kernel
evaluation, i.e. direct local sum and part of translations, while the
data structures alone were sequentially constructed within a single node on CPUs. In
contrast, our approach provides parallel algorithms to build  data
structures not only on the node level, but also at a much finer
granularity within a node, which allows their construction algorithms to be efficiently mapped on SIMD
architectures of GPUs. There are also many other works focusing on a complementary problem: of partitioning the FMM data across multiple processors, such as \cite{Teng1998:JSC}, which shown a provably
and efficient good partition as well as a load balancing algorithm, and
\cite{Barba2010:JNME}, which presented a partition strategy based
on precomputated parameters. 

\subsection{Parallel Hardware}
There is a revolution underway over the past decade or so in the use
of graphics inspired hardware for accelerating general purpose
computation. Since GPUs are attached to the host (CPU) via
PCI-Express bus, processing data on those accelerators requires
data transfer between host and device (GPU). The
on-chip memory are hierarchical and the programming focus is to best use these
hierarchical memories in the threaded model efficiently given the
trade-off between speed and size \cite{Kirk2010:PMP,CudaManual}.



On the many-core GPU accelerators, the data are processed as {\emph
warps}, i.e, a group of threads executing the same instruction at
the same time and thousands of threads are spawned to run in
parallel. Hence, the parallel algorithms presented in this paper are
designed for performance efficiency under this architecture which
favors massively parallel threads and accounts for the cost of memory
access. In this paper, we only focus on the
parallel programming presentations under the NVIDIA GPUs and CUDA, nevertheless our algorithms could be
implemented similarly by using OpenCL \cite{Kirk2010:PMP,CudaOpenCL} or on
different many-core accelerators being introduced such as the AMD APU/GPU or INTEL XEON PHI. 

\subsection{Motivation for Fast Data Structure Algorithms}

Several papers in the literature as we mentioned before
(\cite{Bedorf2012:JCP,Ajmera2008:FPG}, etc.) have been published on fast Kd-tree
and octree data structures that look similar to the spatial data
structures used here, however, they lack the functionality to construct these interaction lists
for the specific neighbor and box query operations, hence
cannot be directly applied in to the FMM framework. The typical way of
computing these data structures is via an $O(N\log N)$ algorithm,
which is built upon  spatial data sorting and is sequentially
implemented on the CPU \cite{Gumerov2008:JCP}. For large dynamic problems (the
particle positions change every time step), this data structure
construction cost would dominate the overall cost by Amdahl's law, especially when the FMM kernel
evaluation is significantly speeded up. Reimplementing the CPU algorithm for the GPU would not achieve the
kind of acceleration we sought. The reason is that the conventional
FMM data structures algorithm employs sorting of large data
and operations such as set-intersection and searching, that require
random access to the global memory, cannot be implemented efficiently
on current GPU architectures.

\section{Single Node Parallel FMM Data Structure Algorithm}
\label{sec3}
The basic FMM data structures in our implementations are based on the
octree \cite[chapter 2]{Samet2005:FMMDS}. At different octree levels,
the unit cube containing all the spatial points is hierarchically divided into
sub-cubes via an octree and each spatial box is assigned a global \textit{Morton index}
\cite{Morton66:GSSS}. Basic concepts and operations on the octree data
structures include: finding neighbor boxes, assigning indices and
finding coordinates of the box center via interleaving/deinterleaving, particle location (box
index) query, etc. Refer to \cite{Gumerov03:DS,Gumerov2004:FMMBook} for details of these basic concepts,
operations and algorithms. 

The algorithm is based on use of occupancy
histograms (i.e., the counts of particles in each box), assigning particles to their grid cells,
and parallel scans \cite{Blelloch89:TC}. A disadvantage of this approach
is the fact that the histogram requires temporary allocation of an array of size $
8^{l_{\max}}$. Nonetheless this algorithm
for GPUs with 4 GB global memory enables of data structures up to a maximum
level $l_{\max}=8$, which is sufficient for many problems. In this
case accelerations up to two orders of magnitude compared to CPU were
achieved. Note that the histogram is only needed at the time of data structure
construction, all the empty box information is skipped in the
final data structure outputs, which are passed to the real FMM kernel
evaluation engine, to achieve both high memory and subsequent summation efficiency.

We would first like to establish
some notation. First of all, all the integers in our implementation,
such as box indices, histograms, are stored as \textit{unsigned int}. We use
\texttt{Src}/\texttt{Recv} to represent source points/receiver points
respectively. We define non-empty
source/receiver boxes as those boxes that contain at least
one source/receiver data point respectively, while empty
source/receiver boxes have no points inside. Note that an empty source
box may contain receiver points and vice versa.

\subsection{Pseudo-Sort Using Fixed-Grid-Method in Linear Time}
\label{SINGLEFMMDS:Sort}
To build the FMM\ data structures, we first need to reorganize the data
points (both source and receiver) into a tree structure
according to their spatial locations, such that at the finest level
each octree box holds at most a prescribed number of points, the {\em cluster size}. By adjusting the cluster size, we
can try to ensure that the costs of the near-field direct sums and the
far-field approximated sums are roughly balanced (or take the same time). Given the cluster
size, we could determine the maximal level $l_{max}$ of the
octree. The data reorganization is realized by a ``Fixed-Grid-Method''
algorithm, in which all data points are rearranged according to their
Morton box indices at level $l_{max}$ but only with linear computation
cost, since the order of data points within a box, which share the
same Morton index, is irrelevant to the algorithm correctness.  We
don't use the word ``sort'' here is because this pseudo-sort is a nondeterministic
algorithm. In our GPU implementation, the final sort order is determined
by the run-time global memory access order of CUDA threads.  

Since we have to pseudo-sort both source and receiver points, we use
the term ``data points'' to refer to both, and denote the
array storing these data points by \texttt{P[]}. Firstly, each data
point \texttt{P[i]} has associated with a 2D vector called
\texttt{sortIdx[i]}, where \texttt{sortIdx[i].x} stores the Morton index
of its box and \texttt{sortIdx[i].y} stores its rank within the
box. Secondly, there is a {\em histogram} array \texttt{Bin[]} allocated for
the boxes at the maximal level. Its $i$th entry
\texttt{Bin[i]} stores the number of data points within the box $i$,
which is computed by the \texttt{atomicAdd()} function in the GPU
implementation. This CUDA function performs a read-modify-write atomic
operation on one 32-bit or 64-bit word residing in global or shared
memory \cite{CudaManual}. Let the number of data points be $M$. Then
the pseudocode to compute \texttt{sortIdx[]} and \texttt{Bin[]} is given in 
Alg.~\ref{algorithm1}.

\begin{algorithm}[tb]
\caption{ \textsc{Parallel-Pseudo-Sort}({\tt P[], M}): an algorithm
to compute the sorted index of each particle using the Fixed-Grid-Method.}
\begin{algorithmic}[1]
  \Require a particle position array {\tt P[]} with length {\tt M}
  \Ensure a 2D index array {\tt sortIdx[]}
  \For {{\tt i=0} to {\tt M-1} {\bf parallel}}
  \State {\tt SortIdx[i].x$\leftarrow$BoxIndex(P[i])}
  \State {\tt atomicAdd[Bin[SortIdx[i].x]]}
  \State {\tt SortIdx[i].y$\leftarrow$Bin[SortIdx[i].x]}
  \EndFor
\end{algorithmic}\label{algorithm1}
\end{algorithm}Although \texttt{atomicAdd()} serializes those threads that access
the same memory address, the parallel performance of
our implementation is good on average. This is because most threads
work on different memory locations at the same time. After this
pseudo-sort, all the data points are copied into a new sorted array
according to their \texttt{sortIdx} and their corresponding bookmark
arrays (a pointer array described in details in
Sec.~\ref{singlenodeDS:interactionlists}), which is used to find the 
data points given a box Morton index,  are constructed. Note that the cost to move data and write the pointer address into the bookmarks are also linear and moving data on the device can take advantage of high GPU memory bandwidth. We 
denote the pseudo-sorted source/receiver point arrays as
\texttt{SortedSrc[]}/\texttt{SortedRecv[]} respectively.

\subsection{Interaction Lists}
\label{singlenodeDS:interactionlists}

\begin{figure}[tb]
\vspace{-0.1in}
\begin{center}
\includegraphics[width=0.4\textwidth]{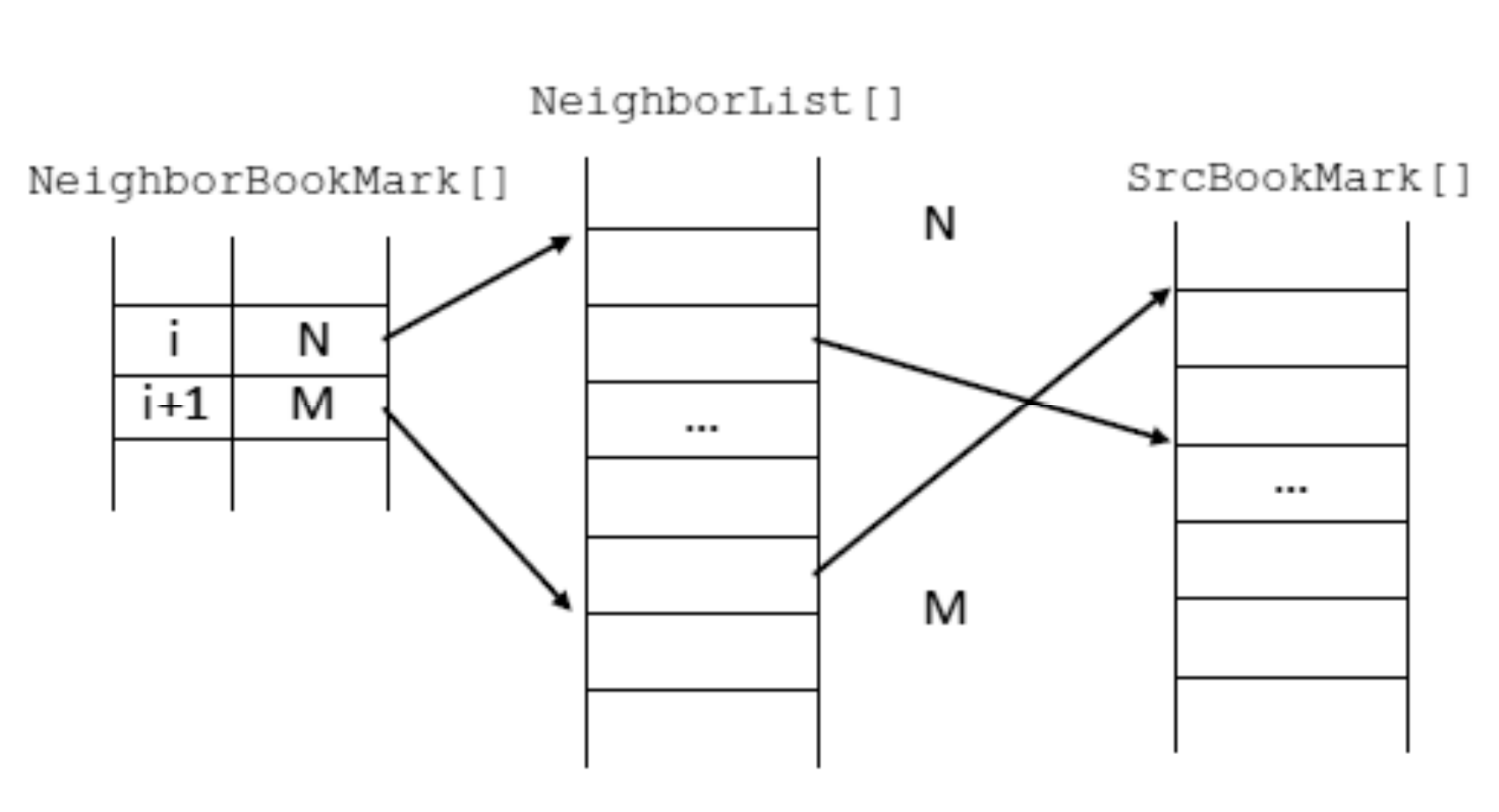}
\end{center}
\caption{The mapping relation among the neighbor bookmark, the neighbor
  list and the source bookmark}\label{fmmNeighbor}
\end{figure}

To access data efficiently, we use several pre-computed arrays, which are also constructed by using parallel algorithms
on the GPU: \texttt{SrcBookmark[]}, \texttt{RecvBookmark[]}, \texttt{NeighborList[]}, \texttt{SrcNonEmptyBoxIndex[]}, and
\texttt{NeighborBookmark[]}. We call these {\em interaction
  lists}. Define $\mathtt{numSrcNEBox}$ and
$\mathtt{numRecvNEBox}$ be the number of non empty source and receiver boxes respectively. We
describe these interaction lists below:

\begin{itemize}
\item \texttt{SrcBookmark[]}: its $i$th entry points to the first sorted
source data point in the $i$th source non-empty box in \texttt{SortedSrc[]}
Its length is $\mathtt{numSrcNEBox+1}$. 

\item \texttt{RecvBookmark[]}: its $j$th entry points to the first sorted
receiver data point in the $j$th receiver non-empty box in \texttt{
SortedRecv[]}. Its length is $\mathtt{numRecvNEBox+1}$.

\item \texttt{SrcNonEmptyBoxIndex[]}: its $i$th entry stores the
  Morton index of the $i$th non-empty source box. Its length is
  $\mathtt{numSrcNEBox}$.

\item \texttt{NeighborBookmark[]}: to perform the local direct sum of
  the $j$th non-empty receiver box, its $E_2$ neighbor information can
  be retrieved from the \texttt{NeighborBookmark[j]}$th$ to the
  \texttt{NeighborBookmark[j+1]-1}$th$ enties in the list
  \texttt{NeighborList[]}.

\item \texttt{NeighborList[]}: given two indices $i$ and $j$ such that
  \texttt{NeighborBookmark[j]} $\leq i<$
  \texttt{NeighborBookm-\ ark[j+1]} and denote
  $k=i-$\texttt{NeighborBookmark[j]}, then \texttt{NeighborList[i]}
  stores the index of the $k$th non-empty source box adjacent to the
  $j$th non-empty receiver box ($E_2$ neighborhood). Here the index
  means the rank of that non-empty source box in the
  \texttt{SrcNonEmpty-\ BoxIndex[]}. See figure~\ref{fmmNeighbor}.

\item \texttt{RecvPermutationIdx[]}: its $i$th entry means the original
position of data point \texttt{SortedRecv[i]} in \texttt{Recv[]} is
\texttt{Rec-\ vPermutationIdx[i]}.
\end{itemize}

\begin{algorithm}[htb]
\caption{\textsc{Access-Source-$E_2$-neighborhood}( {\tt SortedSrc[]}, {\tt SrcBookmark[]}, {\tt
    NeighborList[]}, {\tt NeighborBookmark[]},
 {\tt i}): an
algorithm to extract all the source data within the $E_2$ neighborhood
of the $i$th (non-empty) receiver box.}
\begin{algorithmic}[1]
  \Require the $i$th receiver box and other interaction lists
  \Ensure the source data {\tt tempSrcNei[]} within its $E_2$ neighborhood
  \State {\tt count$\leftarrow$0}
  \State {\tt tempSrcNei$\leftarrow \emptyset$}
  \State {\tt B$\leftarrow$NeighborBookmark[i+1]-1}
  \For {{\tt j=NeighborBookmark[i]} to {\tt B}}
  \State {\tt v$\leftarrow$NeighborList[j]};
  \State {\tt C$\leftarrow$SrcBookmark[v+1]-1}
  \For {{\tt k=SrcBookmark[v]} to {\tt C}}
  \State {\tt tempSrcNei[count++]$\leftarrow$SortedSrc[k]}
  \EndFor
  \EndFor
\end{algorithmic}\label{algorithm2}
\end{algorithm}

Given the bookmark array, the data point can be accessed directly
from the sorted data list. For each
receiver non-empty box, the source data points within its $E_{2}$
neighborhood can be accessed as Alg.~\ref{algorithm2}. The
bookmarks are only kept for non-empty boxes and the neighbor list is
only kept for non-empty neighbors. No information of empty boxes are
passed to the FMM kernel evaluation engine. The last auxiliary array
\texttt{RecvPermutationIdx[]} is used to retrieve the input order of
the original receiver data points.

\subsection{Parallel Data Structure Construction}

\begin{algorithm}[htb]
\caption{\textsc{Get-bookmark-and-box-index}({\tt Bin[]}): an algorithm to
  compute the bookmark and the non-empty box index of source/receiver
  boxes.}
\begin{algorithmic}[1]
  \Require the pseudo-sorted index array {\tt Bin[]} of
  source/receiver boxes
  \Ensure the bookmark array {\tt Bookmark[]} and the Morton index
  array {\tt NonEmptyIdx[]} of source/receiver boxes
  {\Comment \small \emph{array indices depend on implementations}}
  \State perform parallel scan on \texttt{Bin[]} to obtain its prefix
  sum \texttt{ScannedBin[]}
  \For {{\tt i=0} to {\tt Bin[].length-1} {\bf parallel}}
    \If {{\tt Bin[i]>0}}
        \State {\tt Rank[i]$\leftarrow$1}
    \Else
        \State {\tt Rank[i]$\leftarrow$0}
    \EndIf
  \EndFor
  \State perform parallel scan on \texttt{Rank[]} to obtain its prefix
  sum \texttt{ScannedRank[]}
  \State allocate memory for \texttt{Bookmark[]} and
  \texttt{NonEmptyIdx[]}   {\Comment \small \emph{their lengths can be
      derived from \texttt{Bin[]}}}
  \State \texttt{Bookmark[0]$\leftarrow$0} 
  \State \texttt{Bin[-1]$\leftarrow$0}
  \For {{\tt i=0} to
  {\tt Bin[].length-1} {\bf parallel}}
    \If {{\tt Bin[i]>Bin[i-1]}}
        \State {\tt Bookmark[ScannedRank[i]]$\leftarrow$Bin[i]}
        \State {\tt NonEmptyIdx[ScannedRank[i]]$\leftarrow$i}
    \EndIf
  \EndFor
\end{algorithmic}\label{algorithm3}
\end{algorithm}

\begin{algorithm}[htb]
\caption{\textsc{Get-$E_2$-neighbor-list-and-bookmark}(
{\tt Scanned-\ Rank[]}, {\tt RecvNonEmptyBoxIdx[]},
{\tt numNonEmptyRecvBox}): an algorithm to extract the $E_2$
neighborhood for all the (non-empty) receiver boxes.}
\begin{algorithmic}[1]
\Require the {\tt ScannedRank[]} from Alg.~\ref{algorithm3} for
source, the receiver box index array {\tt RecvNonEmptyBoxIdx[]} with
its length {\tt numNonEmptyRecvBox}
\Ensure The $E_2$ neighbor list array {\tt NeighborList[]} (for receiver boxes) and its bookmark
{\tt NeighborBookmark[]}
\State allocate a temporary array \texttt{RecvE2NeiNEBoxIdx[]} to
store neighbor box indices  {\Comment \small \emph{each box can have 27 neighbors at most in 3D}} 
\For {{\tt i=0} to
  {\tt numNonEmptyRecvBox-1} {\bf parallel}}
  \State {\tt $n_i\leftarrow$0}
  \State \texttt{k$\leftarrow$RecvNonEmptyBoxIdx[i]} 
  \For {all its non-empty $E_2$ source neighbor box $j$} 
  \State {\tt NeighborIdx[27i+($n_i$++)]$\leftarrow j$}
  \EndFor
  \State {\tt NumRecvE2NeiNEBox[i]$\leftarrow n_i$}
  \For  {{\tt j=0} to
  {\tt $n_i-1$}}
    \State {\tt RecvE2NeiNEBoxIdx[27i+j]=ScannedRank[Nei-\ ghborIdx[j]-1]}
  \EndFor
\EndFor
  \State perform parallel scan on \texttt{NumRecvE2NeiNEBox[]} to
  obtain its prefix sum \texttt{NeighborBookmark[]}
  \State allocate \texttt{NeighborList[]} {\Comment \small \emph{itself and its length can be
    derived from \texttt {RecvE2NeiNEBoxIdx[]} and
    \texttt{NeighborBookmark[]} respectively}}
  \For {{\tt i=0} to
  {\tt numNonEmptyRecvBox-1} {\bf parallel}}   
    \For {{\tt j=0} to {{\tt NumRecvE2NeiNEBox[i]-1}}}
    \State {\tt count $\leftarrow$ NeighborBookmark[i]+j}
    \State {\tt NeighborList[count]$\leftarrow${\tt
        RecvE2NeiNEBoxIdx[27i+j]}}
     \EndFor
  \EndFor 
\end{algorithmic}\label{algorithm4}
\end{algorithm}

In our implementation, the bookmark for the
source/receiver box is the rank of its first
source/receiver point among all source/receiver points. The bookmark provides a pointer to the  data of any
non-empty box among all boxes without search. A reduction operation is
needed to compute the entries of the bookmark arrays. The highly efficient parallel \textit{prefix sum} (or scan)
\cite{CudaScan} is used in our implementation. Given the
\texttt{Bin[]} obtained from Alg.~\ref{algorithm1}, the
\texttt{Bookmark[]} can be computed by removing the repeated elements
(corresponding to empty boxes) in the prefix sum of \texttt{Bin[]}
using Alg.~\ref{algorithm3}. The same idea can also be used to
address any non-empty source/receiver box among all source/receiver
boxes if we mark non-empty boxes by 1 and empty boxes by 0 and apply
the scan operation. With \texttt{Bookmark[]} and
\texttt{SortIdx[]}, data points are copied to into a new sorted
list. \texttt{SrcNonEmptyBoxIndex[]} is used to construct
\texttt{NeighborBookmark[]} and \texttt{NeighborList[]} in parallel as
Alg.~\ref{algorithm4}: initially a thread computes the $E_{2}$ neighbor
box indices of a non-empty receiver box and checks whether these
source neighbor boxes are empty or not. Then this thread increases the
local non-empty source box count accordingly for its assigned receiver
box and store the neighbor indices temporarily. Finally after another
parallel scan call, the temporary neighbor indices are compressed and written to
\texttt{NeighborList[]}, where the target address is obtained by
reading the non-empty source box index from
\texttt{SrcNonEmptyBoxIndex[]}. Algorithm~\ref{algorithm5}
summarizes all the steps to build the data structures for a single
computing node.

\begin{algorithm}[htb]
\caption{\textsc{Build-fmm-data-structures}({\tt
    Src[], Recv[]}): the single-node algorithm to pseudo-sort data points and
  construct all the needed interaction lists on GPU.}
\begin{algorithmic}[1]
\Require the source/receiver data {\tt Src[]}/{\tt Recv[]}
\Ensure  all interaction lists and the pseudo-sorted source/receiver data {\tt SortedSrc[]}/{\tt
  SortedRecv[]} 
  \State pseudo-sort {\tt Src[]} by Alg.~\ref{algorithm1}
  \State get {\tt SrcNonEmptyBoxIndex[]} and {\tt SrcBookmark[]} by Alg.~\ref{algorithm3}
  \State copy sorted source data points to {\tt SortedSrc[]}
  \State pseudo-sort {\tt Recv[]} by Alg.~\ref{algorithm1}
  \State get {\tt RecvPermutationIdx[]} and {\tt
    RecvBookmark[]}  by Alg.~\ref{algorithm3}
  \State copy sorted receiver data points to {\tt SortedRecv[]};
  \State build {\tt NeighborBookmark[]} and {\tt NeighborList[]}  by
  Alg.~\ref{algorithm2} and Alg.~\ref{algorithm4} 
  \State pass {\tt SortedSrc[]}, {\tt SrcBookmark[]}, {\tt SortedRecv[]}, {\tt RecvBookmark[]}, {\tt RecvPermutationIdx[]},
  {\tt Neighbor-\ Bookmark[]} and {\tt NeighborList[]} to FMM kernel evaluation engine;
  \State free all other allocated device memory;
\end{algorithmic}\label{algorithm5}
\end{algorithm}

All the octree operations needed in Alg.~\ref{algorithm5}, can be found in
\cite{Gumerov03:DS}. By using the {\em interleave} and
{\em deinterleave} operations, we can derive a 3D coordinate for any given
Morton index. Given this 3D vector, we can increase or decrease its
coordinate component to compute its neighbors' 3D
coordinates. Therefore, the algorithms of $E_2$ and $E_4$ neighbor
queries can be easily obtained. Accordingly, they are not
presented as separate algorithms. Note that, given any spatial box, the
computations of its neighbors' coordinates and Morton indices are
independent of other boxes and executed in parallel. 

\subsection{GPU Implementation Considerations}
Basic octree operations, such as box index query, box center query,
box index interleave/deinterleave and parent/children query, and more
complex neighbor query operations, such as $E_2$ and $E_4$ neighbor
index query, are all implemented as inlined CUDA
\texttt{\_\_device\_\_} functions. For efficiency, we minimize the use of global memory and
local memory accessing.  Once input data is loaded into these device
functions, we only use local fast registers, or coalesced local memory
if data can not fit into registers, to store intermediate
results. Moreover, we manually unroll many loops to further optimize
the code. Results shows that even for the
costly computation of $E_4$ neighbors, its total running time can be
neglected in comparison with the kernel evaluation time in the FMM.

\subsection{Complexity}
The complexity of these data structure algorithms is determined by the number of
source points $N$, the number of receiver points $M$ and the maximal
octree level $l_{max}$. Since we use histograms, we can avoid all the searching
operations on the device, which makes our implementation fast and
efficient. However, there is a memory consumption trade-off for the
processing speed since the size of histogram increases exponentially
as $l_{max}$. For the bucket
sort Alg.~\ref{algorithm1}, its complexity is linear
$O(N+M)$. All other algorithms are related to the octree boxes, which
total number is $N_{box}=2^{3l}=8^l$. Since we use the canonical scan
algorithm, Alg.~\ref{algorithm2} to Alg.~\ref{algorithm4} are in
$O(N_{box}\log N_{box}+N+M)\sim O(l8^l+N+M)$. If we interpret the
maximal level $l_{max}$ as a prescribed constant, then our parallel data
structure construction Alg.~\ref{algorithm5} for single node is
linear with respect to particle size, i.e. in $O(N+M)$.

\section{FMM Data Structures on Multiple Nodes}
\label{sec4}

\begin{figure}[t]
\begin{center}
\hspace*{-0.08in}
\includegraphics[width=0.5\textwidth]{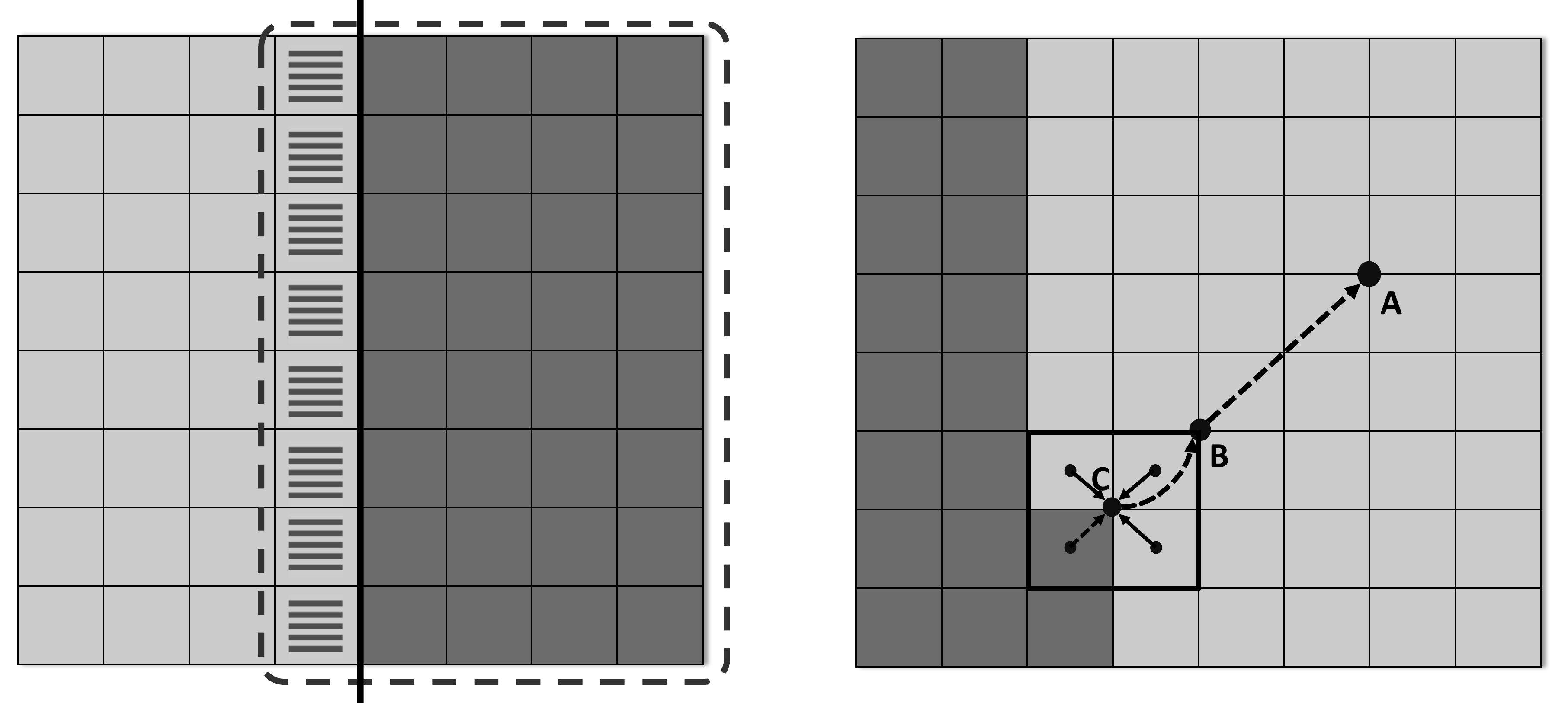}
\end{center}
\caption{Problems in distributing the FMM across two nodes. Left: lightly-shaded  boxes are on node 1 (Partition I) and darkly shaded boxes are on node 2 (Partition II). The thick line indicates the partition boundary line and the dash line shows the source points in both partitions needed by Partition II. The hashed boxes are in Partition I but they have also to be included in Partition II to compute the local direct sum. Right: light boxes belong to Partition I and dark boxes belong to Partition II. The multipole coefficients of the box with thick lines (center at C) is incomplete due to one child box on another node. Hence its parents (B and A) in the tree up to the minimal level are all incomplete.}\label{partitionProblem}
\end{figure}

Since all the data structures are constructed based on the
locations of source and receiver data points, there are two main issues on
the multiple nodes. First, on multiple nodes using the algorithm of
\cite{Qi11:SC11,Qi12:HPCC12}, only receiver data points are mutual exclusively distributed. Source points which are in the
halo regions (boundary layers of partitions) have to be distributed on several nodes, because
of the direct sum region overlap. Hence on each node, the source data
points for direct sum and translation are no longer the same.
When we build the translation data structure, these
repeated source data should be guaranteed to translate only once among all the
nodes. Second, since any translation
stencils may require coefficients from many source boxes which are on other
nodes, there will be many translation
coefficient communications among different nodes. Moreover, due to the partition,
from a certain level on, the octree box coefficients on a single node might
be incomplete up to the root of the octree. This is because
if one of any box's children is distributed on one or several
different nodes, all its ancestors coefficients are incomplete. In
Fig.~\ref{partitionProblem}, we show such an example. Finally, all
information related to these boxes are stored in a compressed way
because we skip all empty boxes. Therefore it is a non-trivial task to
fetch the coefficients of any box efficiently since many searching and rearranging
data operations are needed. Good partition and data communication algorithms are
crucial to reduce the communication overhead in terms of both data
transfer size and data packing time.

\subsection{Global Data Structure and Partitioning}
In \cite{Qi11:SC11} FMM algorithms for the heterogeneous CPU-GPU
architecture were explored and it was concluded that a good strategy
is to distribute the FMM computation components
between CPUs and GPUs: expensive but highly parallizable particle
related computations (direct sums) are assigned to the GPU, while the
extensive and complex space box related computations (translations) are
assigned to CPU. This way one can take the best advantages of both CPU and GPU hardware
architecture, and we design our data structures for this mapping. 



The split of the global octree can be viewed as a forest of
$K$ trees with roots at level 2 and leaves at level $l_{max}$. In the
case of more or less uniform data distributions and number of nodes
less than $K \leq 64$ (for the octree), each node may handle one or
several trees. If the number of nodes are more than 64 and/or data
distributions are substantially non-uniform, partitioning based on the
work load balance can be performed by splitting the trees at levels
$>$ 2. Such partitioning can be thought as breaking of some edges of
the initial graph. This increases the number of the trees in the
forest, and each tree may have a root at an arbitrary level $l = 2,
..., l_{max}$. Each node then takes care for computations related
to one or several trees. At this point we assume that there exists
some work load balancing algorithm which provides an efficient
partitioning. At this point we also do not put any constraint to
interaction between the receiver and source trees, so formally this
can be considered as two independent partitions.

For presentation purposes, we define two important concepts although
they are related to each other in our
implementation:
\begin{itemize} 
\item {\bf partition level} $l_{par}$:
  at this level, the whole space are partitioned among different
  computing nodes. On a local node, all the subtrees at this level or
  below are totally complete, i.e., no box at level $\geq l_{par}$ is
  on other nodes.
  \item {\bf critical level} $l_{crit}$: at this level,
  all the box coefficients are broadcasted such that all boxes at level
  $\leq l_{crit}$ can be treated as local boxes, i.e., all the box
  coefficients are complete after broadcasting. In our implementation,
  $l_{crit}=\max{(l_{par}-1,2)}$.
\end{itemize}Normally the number of computing nodes in current high
performance clusters are in the order of hundreds or even thousands, that
is considered much smaller than the number of spatial boxes of the global
octree. Hence, the partition level is usually quite low, such as
$l_{par}=2,3,4$. Hence broadcasting the coefficients at the critical
level $l_{crit}$ only requires a small amount of data with
neglectable communication overhead given the major cost of kernel evaluations.


\begin{figure}[t]
\begin{center} \hspace*{-0.1in}
\includegraphics[width=0.3\textwidth]{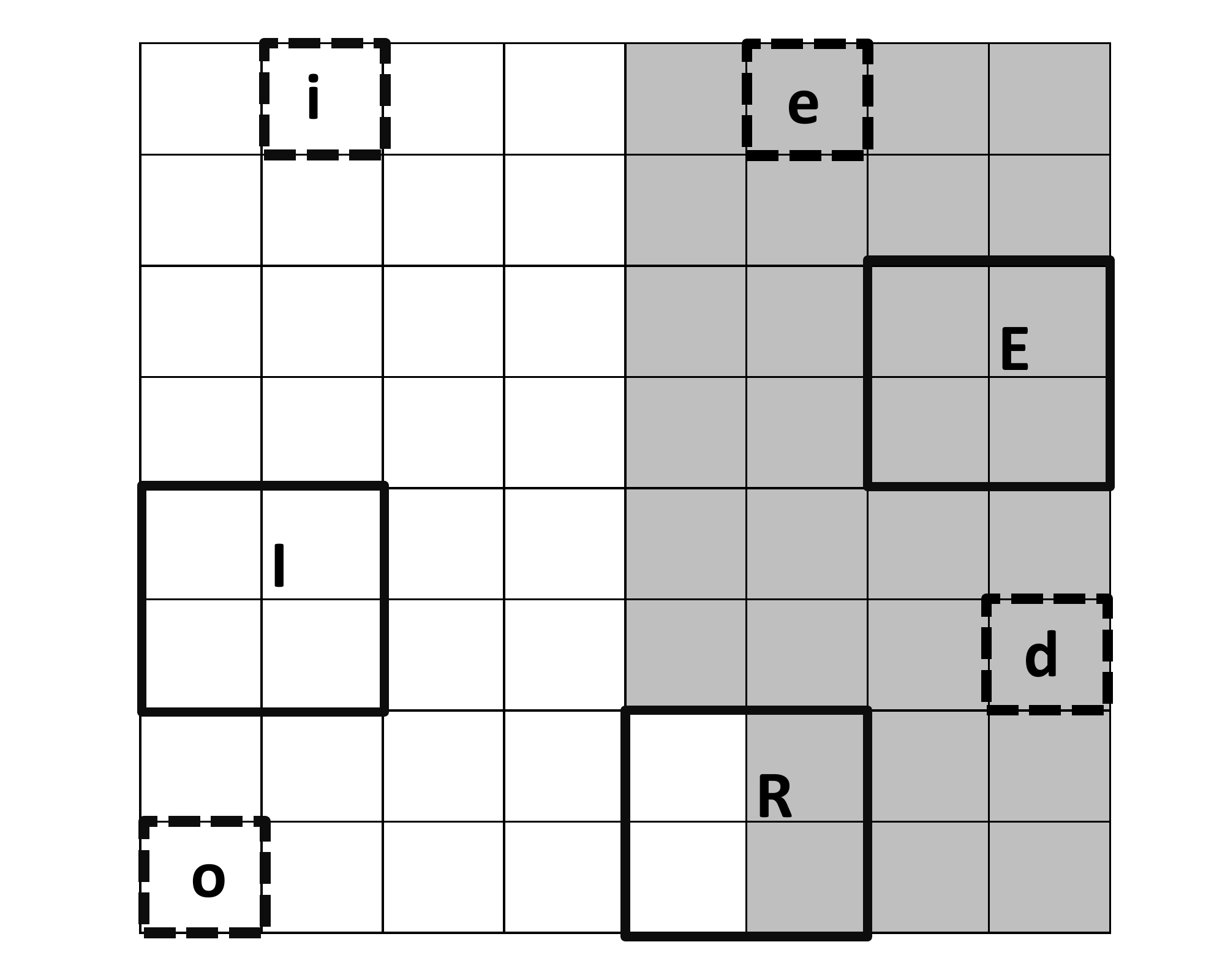}
\end{center}
\caption{An example of source box types. White boxes are Partition I and gray boxes
are Partition II. The partition level is 3 and the critical level is
2. Solid line boxes correspond to level 2 and dash line boxes
correspond to level 3. At partition II, box $\mathbf{e}$ and $\mathbf{E}$ are {\em export boxes}. Boxe $\mathbf{i}$ and $\mathbf{I}$ are {\em import boxes}. Box $\mathbf{R}$ is a {\em root box}. Box $\mathbf{d}$ is a {\em domestic box}. Box $\mathbf{o}$ is an {\em other box}.}\label{boxType}
\end{figure}

To better organize data communication, on each node, we classify all the source
boxes into five categories in an array \texttt{SrcNonEmptyBoxType[]}. From the finest level
$l_{max}$ to 2, we list all the global source box Morton
indices in an increasing order. Note that some local empty source boxes
might be in the list since this box may contain source points which
are located on other nodes and this global source boxes information is
obtained from the initial global octree construction. Given those box
types, we could determine which boxes need to import or export their 
$M$-coefficients. For any node, say $J$, these five box types are (see
Fig.~\ref{boxType}):
\begin{enumerate}
\item {\bf Domestic Boxes}: The box and all its children are on
  $J$. All domestic boxes are organized in trees with roots located at
  level 1. All domestic boxes are located at levels from $l_{max}$ to
2. The roots of domestic boxes at level 1 are not domestic boxes (no
data is computed for such boxes).

\item {\bf Export Boxes}: These boxes need to send data to other
nodes. At $l_{crit}$, the M-data of export boxes may be incomplete. At
level $> l_{crit}$, all export boxes are domestic boxes of $J$ and their M-data are complete.

\item {\bf Import Boxes}: Their data are produced by other computing
nodes for importing to $J$. At $l_{crit}$, the M-data of import
boxes may be incomplete. At level $> l_{crit}$, all import
boxes are domestic boxes of nodes other than $J$ and their M-data are
complete there.

\item {\bf Root Boxes}: These are boxes at critical level, which need
  to be both exported and imported. For level $>l_{crit}$ there is no root box.

\item {\bf Other Boxes}: Boxes which are included in the data
structure but do not belong to any of the above types, e.g. all boxes
of level 1, and any other box, which for some reason is passed to the
computing node (such boxes are considered to be empty and are skipped in
computation, so that affects only the memory and amount of data
transferred between the nodes).
\end{enumerate}

Note that there are no import or export boxes at levels from $l_{crit}-1$ to
2. All boxes at these levels are either domestic boxes or other boxes
after the broadcast and summation of incomplete $M$-data at
$l_{crit}$. We only need compute M-data and box types from level
$l_{max}$ to $l_{crit}$ and exchange the information at
$l_{crit}$. After that we compute the $M$-data for all the domestic
boxes up to level 2 then produces $L$-data for all receiver boxes at
level $l_{max}$ handled by the computing node.

\subsection{FMM Algorithm on Multiple Nodes}
Our multiple node algorithm involves three main parts: \begin{enumerate}
\item Global source and receiver data partition: the partition should keep work balance among all the computing nodes.
\item Single node evaluation: a single node performs the translations upward/downward, compute the export and import data and the local summations.
\item Multiple node data exchange: The data communication manager collects and distributes the data from/to all the computing nodes accordingly.
\end{enumerate} Parts (2) and (3) are mutually inclusive because the
translations on a single node require the missing data from other
nodes while the data communication requires import and export
information from each computing node. Part (1) depends on the
application. For dynamic problems, the FMM evaluation is performed for
every time step and the data distribution can be derived from the
previous time step. It is very likely that all the nearby data are
stored on the same node, in which case the partition to keep work
balance only requires a small amount of communication. For
problems only performing a single FMM evaluation, the data appear on
each node might be dependent on some geometric properties but it is
also possible that the initial data on each node is random, in
which case a large amount of the inter-node
communications is inevitable. In our implementation, we assume the worst case that
all the data on each node are random.


\subsubsection{Global Partition}
The architecture of our computing system, and perhaps of most current and near future
systems, is heterogeneous. Each node has several multicore
CPUs and one or two many-core GPUs. While the CPU cores on the same node share
the main memory, each GPU has its own dedicated device memory,
connected to host via PCI-Express bus. To perform computations on
GPUs, the data for a single node have to be divided again for each
GPU. As mentioned before, we perform direct sums on GPUs and FMM
translation on CPUs, hence we need two level partition: divide the
data for nodes first (translation) then further divide data cdof each node for each
GPU (direct sum). Given the prescribed cluster size, we construct the global octree
then split it, i.e., the partition of all the data is performed
by boxes but not by particles.

Assume the same number $g$ of GPUs on each computing node, then we
implement this two level partition as follow: we assign a unique
global ID $(ig+j)$ to the $j$th GPU on the node $i$ and compute our finer
partition with respect to those GPUs. From $l_{par}=2$, our algorithm
tries to distribute all the boxes at level $l_{par}$ among GPUs such
that the amount of receiver points satisfy the prescribed balance
conditions. It increases the $l_{par}$ by 1 until the work load
balance is roughly achieved. Once this finer partition is done, we
automatically obtain a balanced coarse partition with respect to
computing nodes (this is because each node has the same number of
GPUs). To identify all box locations, we use an auxiliary array
{\tt BoxProcId[]}, in which $i$th entry stores the GPU ID where the
$i$th box at $l_{par}$ resides in. Dividing {\tt BoxProcId[i]} by $g$,
we can obtain the node ID where the $i$th box resides in. Note that,
for any given box at any level, we can easily answer its location
query by shifting that box's Morton index and examining {\tt
  BoxProcId[]}, i.e. by checking its ancestor/children's
location. Initially, we use GPUs and Alg.~\ref{algorithm1} to
pre-process data, i.e. to get the number of receiver points in
each box. Then all nodes send their {\tt Bin[]} array to the master
node. The master node then computes the balanced partition
and derives the value of $l_{par}$ such that each GPU is assigned several
spatial boxes at $l_{par}$ with consecutive Morton indices. Finally,
the master node broadcasts this partition information to all the nodes
and each nodes distributes its own source and receiver data based on
the partition to others. 

\subsubsection{The FMM Algorithm on Multiple Nodes}\label{multinodefmm}
Assume that the balanced global data partition and distributed data
are available. On one hand, the data structure constructions of local neighbor
interaction lists for direct sum are the same as section
\ref{sec3}. On the other hand, the
data structures of translations are for the coarse partition (with
respect to node), hence they need to be recomputed by merging the
octree data structures obtained from
multiple GPUs on the same node. The merging steps are conducted as follows
\begin{enumerate}
\item  Extract all the global source box
  information across all the computing nodes: after all GPU calls the
  data structure construction call of section~\ref{sec3}, each node
  collects these non-empty source box indices from all its GPU, merges
  to one list and send to the master node. Then the master node merges
  all the lists to one global non-empty source box array and
  broadcasts to all nodes.
\item Extract the local receiver box information for each node: each
  node collects these non-empty receiver box indices from all its GPU,
  merges to one list. Because each GPU deals with consecutive receiver
  boxes, this merging is actually equivalent to copy operations.
\end{enumerate}Once these two box index arrays are available, we can construct the interaction lists for translation
stencils, in parallel on GPU. Except the source box type, the
algorithm for which will be described later, all other needed information
arrays, such as neighbor lists/bookmark etc, can be obtained by using
the algorithms in section \ref{sec3}. Each node $J$ then executes the following translation
algorithm:
\begin{enumerate}
\item {\bf Upward translation pass}:\begin{enumerate}
\item Get M-data of all domestic source boxes at $l_{max}$ from GPU
  global memory.
\item Produce M-data for all domestic source boxes at levels $l =
l_{max}-1,\ldots, \max(2, l_{crit})$.
\item Pack export M-data, the import and export box indices of all
levels. Then send them to the data exchange manager.
\item The master node, which is also the manager, collects data. For
the incomplete root box M-data from different nodes, it sums them
together to get the complete M-data. Then according to each node's
export/import box indices, it packs the corresponding M-data then
sends to them.
\item Receive import M-data of all levels from the data exchange
manager.
\item If $l_{crit} > 2$, consolidate $S$-data for root domestic boxes
at level $l_{crit}$. If $l_{crit} > 3$, produce M-data for all domestic source boxes
at levels $l = l_{crit} - 1,\ldots, 2$.
\end{enumerate}
\item {\bf Downward translation pass}:\begin{enumerate}
\item Produce L-data for all receiver boxes at levels $l = 2,\ldots,
l_{max}$.
\item Output L-data for all receiver boxes at level $l_{max}$.
\item Redistribute the L-data among its own GPUs.
\item Each GPU finally consolidates the L-data, add the local sums to the dense sums
and copy them back to the host according to the original inputting
receiver's order.
\end{enumerate}
\end{enumerate} When each node outputs its import box index array, they are listed in an increasing
order from $l_{max}$ to $l_{crit}$. The manager processes the
requesting import box index array one after another. Given the $i$th
requested source box index \texttt{ImportSrcIdx[i]}, the manager first
figures out its level $l_*$. If $l_*<l_{crit}$, the manager derives
\texttt{ImportSrcIdx[i]}'s ancestor in the partition level and check
the array \texttt{BoxProcId[]} to find which node it belongs to. If
$l_*==l_{crit}$, the manager will check all its children's node
address ($l_*$ can not exceed $l_{crit}$ since all the boxes above the
critical level are marked as domestic box). Once the manager
identifies the node ID, where that box belongs to or its children
belong to, it searches the export box index array from that node for
\texttt{ImportSrcIdx[i]} at level $l_*$ then makes an copy of the corresponding
$M$-data in the sending buffer.

Even though each node handles a large number of spatial boxes, the
amount information exchanged with the manager is actually small since
only boxes on the partition boundary layers need to be transferred back and forth. In our current implementation,
the manager is responsible for all the collecting and redistributing
$M$-data work, which involves searching operations, the total run time
is still smaller by comparing with the communication scheme of
\cite{Qi11:SC11}, where all the box's $L$-data ($O(p^28^{l})$ in the
case of uniform distribution) at the finest level $l_{max}$ are
broadcast from the master node.

\subsubsection{Source Box Type}
The type of a source box $\mathbf{k}$ is determined by the M2M and M2L
translation because its children or neighbors might be missing due to data
partition and have to be requested from other nodes. However, once
the parent box M-data is complete, the L2L translations for its
children are always complete. So we can summarize the key idea of Alg.~\ref{algorithm6}, which computes
the type of each box, as follows:
\begin{itemize}
\item At the critical level, we need all boxes to perform upward M2M
translations. If one child is on a node other than $J$,
its M-data is either incomplete or missing, hence we mark it an
import box. We also check its neighbors required by M2L
translation stencil. If any neighbor is not on $J$, then the
M-data of these two boxes have to be exchanged.
\item For any box at the partition level or deeper levels, if this box
is not on $J$, then it is irrelevant to this node, in which
case it is marked as other box. Otherwise we check all its neighbors
required by M2L translations. Again if any neighbor is not on
$J$, these two boxes' M-data have to be exchanged.
\end{itemize}

\begin{algorithm}[!h]
\caption{\textsc{Get-source-box-type}(\texttt{BoxIndex[]},\texttt{ParInfo},$J$): the
  algorithm to compute the source box types on the Node $J$ given the
  partition information}
\begin{algorithmic}[1] 
\Require a source box index \texttt{BoxIndex[i]}$=\mathbf{k}$ at level
$l$, the partition information and the node ID $J$
\Ensure \texttt{BoxType[i]}
\State \texttt{isOnNode}$\leftarrow${\tt isImportExport}$\leftarrow${\tt
  isExport}$\leftarrow${\tt FALSE}
\If {$l$\texttt{<}$l_{crit}$}
\State \texttt{BoxType[i]$\leftarrow$DOMESTIC}
\ElsIf{$l$\texttt{=}$l_{crit}$}
\For{any $\mathbf{k}$'s child $\mathbf{c}_i$ at partition level}
\If {$\mathbf{c}_i$ is not on $J$}
\State \texttt{isImportExport$\leftarrow$TRUE}
\Else
\State \texttt{isOnNode$\leftarrow$TRUE}
\EndIf
\EndFor
\If {\texttt{isOnNode=FALSE}} 
\State \texttt{BoxType[i]$\leftarrow$IMPORT}
\Else \For{any $\mathbf{k}$'s neighbor of M2L translation $\mathbf{n}_i$}
\If {one of $\mathbf{n}_i$'s children at $l_{crit}$ is not on $J$}
\State {\tt isExport$\leftarrow$TRUE} {\Comment \small \it update the type of a different box }
\State $\mathbf{n}_i$'s box type $\leftarrow$\texttt{IMPORT}
\EndIf
\EndFor
\EndIf
\Else 
\If { $\mathbf{k}$'s ancestor at $l_{crit}$ is not on $J$}
\State \texttt{BoxType[i]$\leftarrow$OTHERS}
\Else \For{any $\mathbf{k}$'s neighbor of M2L translation $\mathbf{n}_i$}
\If {the ancestor of $\mathbf{n}_i$ at $l_{crit}$ is not on $J$}
\State {\tt isExport$\leftarrow$TRUE}{\Comment \small \it update the type of a different box }
\State $\mathbf{n}_i$'s box type $\leftarrow$\texttt{IMPORT}
\EndIf
\EndFor
\EndIf
\EndIf

\State synchronize all threads
\If{{\tt isImportExport=TRUE}}
\State \texttt{BoxType[i]$\leftarrow$ROOT}
\ElsIf{{\tt isExport=TRUE}}
\State \texttt{BoxType[i]$\leftarrow$EXPORT}
\Else \State \texttt{BoxType[i]$\leftarrow$DOMESTIC}
\EndIf
\end{algorithmic}\label{algorithm6}
\end{algorithm}

We compute all box types in parallel on the GPU. For each
level from $l_{max}$ to 2, a group of threads on the node $J$ are
spawned and each thread is assigned by one source box index at that
level. After calling Alg.~\ref{algorithm6}, all these threads
have to be synchronized before the final box type assignment in order
to guarantee no race conditions. Note that some ``if-then" conditions
in Alg.~\ref{algorithm6} can be replaced by {\tt OR} operations
so that thread ``divergent branches" can be reduced.

\subsection{Complexity}

Lets assume that we have $P$ computing nodes and each node has $g$
GPUs. We only count non-empty boxes here and all symbols are used as
follows: $B^{src}_{i,l}$ and $B^{recv}_{i,l}$ are the numbers of local
source and receiver boxes at level $l$ on the $i$th node respectively;
$B^{src}_{all,l}$ is the number of global source boxes at level $l$;
$N_i$ and $M_i$ are the numbers of source and receiver points on the
$i$th node respectively; $N$ and $M$ are the total numbers across all
the nodes. 

First, we estimate the running time of our global
partition given the worst case (the initial data are totally random):
\begin{equation}
T_1=a_0(N_{i}+M_{i})+a_1B^{recv}_{i,l_{max}}+a_2B^{recv}_{all,l_{max}}+a_3B^{recv}_{all,l_{max}}\log P+a_4(N+M). \label{eqn1}
\end{equation} Each term within the equation~(\ref{eqn1}) is described
in table~\ref{table1}. Note that, moving $O(N+M)$ data points is
inevitable if the initial data on each node are random, however in
many applications, this communication cost can be avoided or reduced
substantially if this initial distribution is known. Also there are many
publications on this initial partition (tree generation) and optimized
communication in the literature such as \cite{Teng1998:JSC,Ying2003:SC03,Barba2010:JNME}, which can be used
for different applications accordingly.

\begin{table}[htb]
\begin{center}
\begin{tabular}{|r|p{2.3in}|}
  \hline
  Term & Description \\ \hline
  $a_0(N_{i}+M_{i})$ & each node derives the local
  box Morton indices for its source and receiver points and sends them \\
  \hline
  $a_1B^{recv}_{i,l_{max}}$ & each node sends its receiver box
  indices at the finest level to the master node \\ \hline
  $a_2B^{recv}_{all,l_{max}}$ &  the master node collects all the
  receiver box indices and builds the global indices \\ \hline
  $a_3B^{recv}_{all,l_{max}}\log P$ & the master node broadcasts the
  global receiver box indices and partition \\ \hline
  $a_4(N+M)$ & all nodes exchange source and receiver data according to
  the partition (including a node-wide synchronization) \\
  \hline
\end{tabular}\caption{Description of
  equitation~(\ref{eqn1})}\label{table1}
\end{center}
\end{table}

Second, Alg.~\ref{algorithm6} examines the occupancy status of
each source box's $E_4$ neighbors. Note that the number of $E_4$
neighbors for any octree box is upper bounded by 27 and for each
neighbor such check operation is in constant time. Hence the cost to
compute all source box types can be estimated as:
\begin{equation}
T_2=b_0B^{src}_{all,l_{max}}+b_1B^{src}_{all,l_{max}-1}+\ldots+b_{l_{max}-2}B^{src}_{all,2}
\leq b_*B^{src}_{all,l_{max}}. \label{eqn2}
\end{equation}

Finally, let's denote
$B^{src}_{all}=\sum_{j=2}^{l_{max}}B^{src}_{all,j}$. There is no clean
model to estimate the number of export/import boxes. However, the
boundary of each partition is nothing but a surface of some 3D object. Given the uniform
distribution case, in which we can simplify the model, it is
reasonable to estimate the exchanged box number as
$\mu2^{2l_{max}}=\mu4^{l_{max}}$ with some constant $\mu$ for all the
nodes (\cite{Lashuk2009:SC09} estimate this number as
$O((B^{recv}_{i,l_{max}})^{2/3})$, which is similar as
ours). Therefore, at the critical level $l_{crit}$, the exchanging
data cost can be estimated as:
\begin{equation}
T_3=c_0B^{src}_{all}+c_1\mu4^{l_{max}}p^2+c_2\mu4^{l_{max}}\log
(\mu4^{l_{max}})+ c_38^{l_{crit}}p^2. \label{eqn3}
\end{equation}
\begin{table}[htb]
\begin{tabular}{|c|p{2.2in}|}
  \hline
  Term & Description \\ \hline
  $c_0B^{src}_{all}$ & each node examines its source box types and
  extracts import and export source box indices \\ \hline
  $c_1\mu4^{l_{max}}p^2$ & each node sends the export box's $M$-data
  to the master node \\ \hline
  $c_2\mu4^{l_{max}}\log (\mu4^{l_{max}})$ & the master node addresses
  all the requested import box indices of each node\\ \hline
  $c_38^{l_{crit}}p^2$ & the master node packs all the import $M$-data
  and sends to other corresponding nodes \\ \hline
\end{tabular}\caption{Description of equitation~(\ref{eqn3})}\label{table2}
\end{table}Each term in the equation~(\ref{eqn3}) is explained in
table~\ref{table2}. By combining some constant coefficients, we can
further simplify $T_3$ as
\begin{equation}
\begin{tabular}{rcl}
$T_3$&=&$c_0B^{src}_{all}+c_1\mu4^{l_{max}}p^2+c_2\mu l_{max}4^{l_{max}}\log (4\mu)+ c_38^{l_{crit}}p^2$, \\
     &=&$d_0B^{src}_{all}+d_1l_{max}4^{l_{max}}+ d_2(4^{l_{max}}+d_38^{l_{crit}})p^2$.
\end{tabular}
\label{eqn4}
\end{equation}Because those three parts are executed sequentially, $T$, the
total cost of communications and data structures, can be obtained as
\begin{equation}
\begin{array}{rl}
T=&T_1+T_2+T_3=T_{gpu}+T_{cpu}+T_{comm},\textrm{ where}\\
T_{gpu}\leq&a_0M_i+a_1B^{recv}_{i,l_{max}}+b_*B^{src}_{all,l_{max}}\\ 
=&a_0 M/P+a_1 B^{recv}_{all,l_{max}}/P+b_*B^{src}_{all,l_{max}},\\
T_{cpu}=&d_0B^{src}_{all}+d_1l_{max}4^{l_{max}},\\
T_{comm}=&d_2(4^{l_{max}}+d_38^{l_{crit}})p^2+a_2B^{recv}_{all,l_{max}}+a_3B^{recv}_{all,l_{max}}\log
P \\ 
&+a_4(N+M).
\end{array}
\label{eqn6}
\end{equation}

The ideal algorithm should have all costs proportional to $1/P$. In our case, only
the particle related terms but not all the box related terms are
amortized among all the computing nodes. However, in practical,
$l_{max}$ can not be very large which is viewed as a constant in
most cases. More sophisticated schemes can be used so that these
theoretical non-scalable terms can be amortized among all the nodes.
However, given our parallel implementation, the constant
coefficient for each term is small even when the box
number is large. Hence $T_{gpu}$ and $T_{cpu}$ could be negligible compared with
the kernel time.
\begin{figure*}[!htb]
\begin{center}
\includegraphics[width=0.9\textwidth]{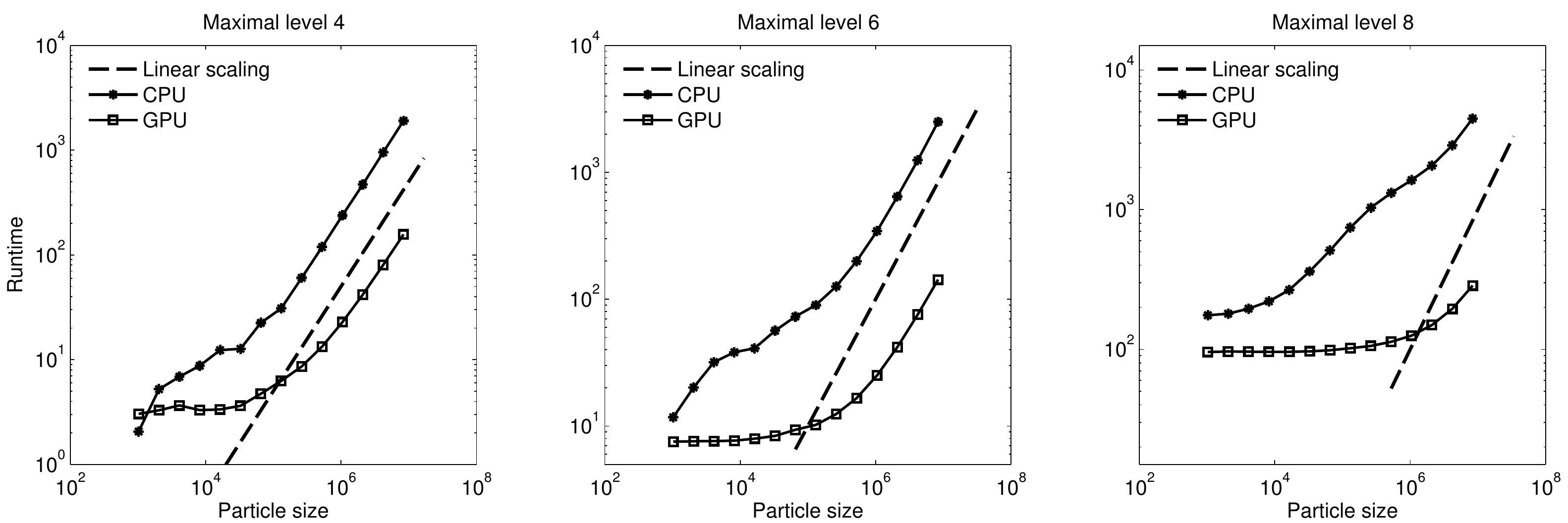}
\end{center}
\caption{The data structure construction time for non-uniform
  distribution on a single node using one GPU: all the source and receiver are distributed on a
  sphere. Each sub-figure corresponds to a maximal level setting.}\label{spheredstime}
\end{figure*}

As for the communication part, the real killing
communication comes from $a_4(N+M)$ since we target on
billion scale problems. Exchanging all these particle data requires
much more time than the real kernel evaluations. However, this term is
not encountered usually because it is obtained from the worst
case, that is the totally random distribution. In most application,
based on the physical or geometric properties, 
this initial data distribution can be configured such that only small
amount of particle communication is needed. Moreover, as mentioned before, we
could use the parallel partition methods in literature to minimize
this cost. All other terms in $T_{comm}$ are still scalable since they
are determined by the number of non-empty spatial boxes ($\ll N$ or
$M$) and these costs are much less than the kernel evaluation time.

\section{Experimental Results}
\label{sec5}

\subsection{Single Node Algorithm Test}

To test the data structure performance for a single node, we fix the
problem size to 1 million and use the uniform distributed source
and receiver. Here the source and receiver points are different. 
The computation hardware used here are: NVIDIA GTX480 GPU and
Intel Xeon X5560 quad-core CPU running at 2.8GHz. 

\begin{table}[htb]
\centering
\begin{tabular}{|c|c|c|c|}
\hline
$l_{\max}$ & CPU (ms) & Improved CPU (ms) & GPU (ms) \\ \hline
3 & 1293 & 223 & 7.7 \\
4 & 1387 & 272 & 13.9 \\
5 & 2137 & 431 & 13.0 \\
6 & 8973 & 1808 & 34.6 \\
7 & 30652 & 6789 & 70.8 \\
8 & 58773 & 7783 & 124.9 \\ \hline
\end{tabular}
\caption{The time comparison of FMM data structure computation for $2^{20}$ uniform randomly distributed
source and receiver particles using our original CPU $O(N\log N)$
algorithm, the improved $O(N)$ algorithm on a single CPU core, and its
GPU accelerated version.}
\label{dsTime}
\end{table}

We firstly test the performance on the uniformly distributed data in a
unit cube. Note that, this would be most time consuming case since
almost all the spacial boxes are non-empty. The timing 
results are summarized in Table~\ref{dsTime}, in which the octree depth was varied in the range $l_{\max}=3,...,8$. Column 2
shows the wall clock time for a standard algorithm, which uses sorting
and hierarchical neighbor search using set intersection (the neighbors
were found in the parent neighborhood domain subdivided to the children
level). Column 3 shows the wall clock time for the present algorithm
on the CPU. It is seen that our algorithm is several times faster.
Comparison of the GPU and CPU times for the same algorithm show further
acceleration in the range 20-100. 

In the second experiment, we generate all the source and receiver data
on a sphere surface and test how the algorithm scales and the
performance gain. In Figure~\ref{spheredstime}, we show both the CPU and GPU
time across the number of data points, which ranges from 1024 up to 8
millions, for different octree maximal levels. As the $l_{max}$
increases, there are more spatial boxes occupied which {\it super
  linearly} increases the overall costs. However, once the number
boxes become relative stable, i.e., increasing the number of particles
only changes the number of spatial boxes a little bit, the overall
cost increases linearly. This is because that all the boxes related constructions is more or
less the same as a constant and the particle related computation, such
as bit interleaving and the fixed-grid-method pseudo-sort that linearly scales as the amount
of particle data, now dominates the overall
costs. In a whole, for this non-uniform distribution, our data
structure algorithms also demonstrate their linear complexity and our
fast parallel implementations can achieve 15-20 times speed-ups
against the CPU performance.   

As a conclusion of these tests on a single node, the FMM data structure step is
reduced to a small part of the computation time again, which provides
substantial overhead reductions and makes our algorithm suitable to
solve dynamic problems.


\subsection{Multiple Node Algorithm Test} 

\begin{figure}[t]
\begin{center}
\includegraphics[width=0.42\textwidth]{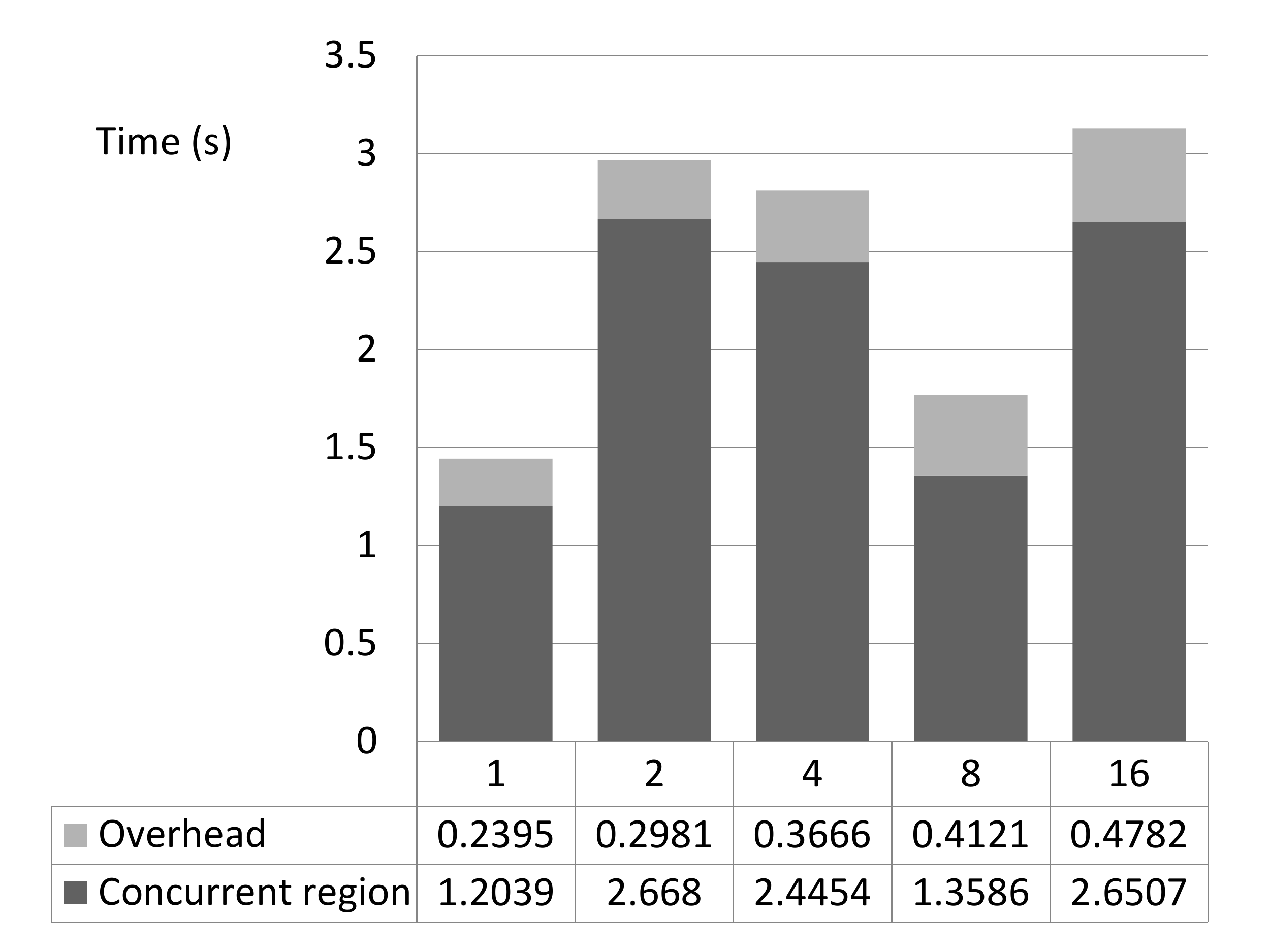}
\end{center}
\caption{The CPU/GPU concurrent region time vs the overhead (data
transfer between the nodes and CPU/GPU sequential region) of the FMM
algorithm for 2 GPUs per node. The testing case size increases
proportionally to the number of nodes (8M particles per node).}\label{overheadComp}
\end{figure}

\begin{figure}[hbt]
\begin{center}
\hspace*{-0.2in}
\includegraphics[width=0.5\textwidth]{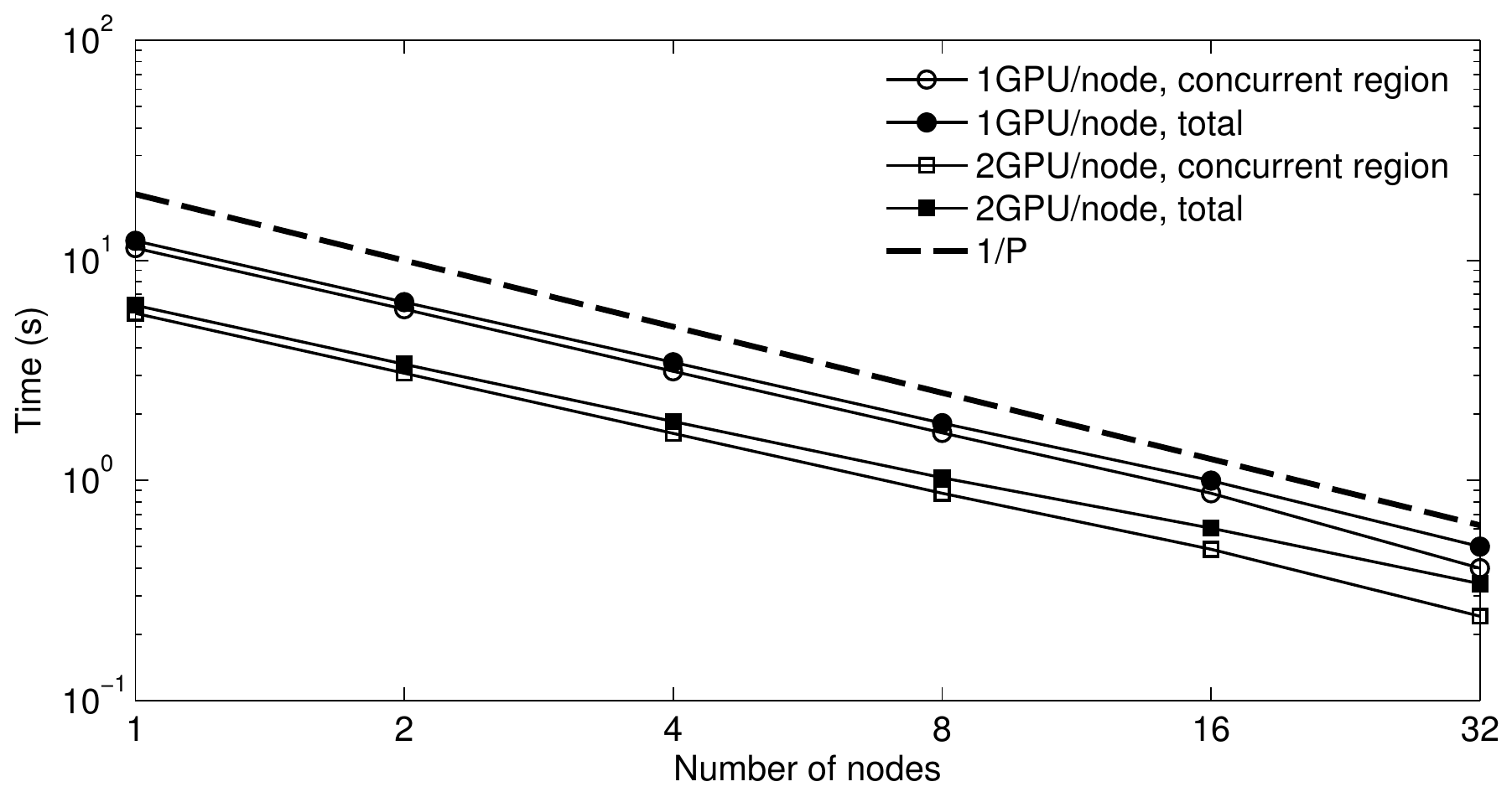}
\end{center}
\caption{The results of the strong scalability test for 1 and 2 GPUs
per node of the testing cluster. The thick dashed line shows perfect
scalability $t=O\left( 1/P\right) $. The problem size is fixed to be 16M.}
\label{strongScale}
\end{figure}


\begin{figure}[htb]
\begin{center}
\hspace*{-0.07in}
\includegraphics[width=0.5\textwidth]{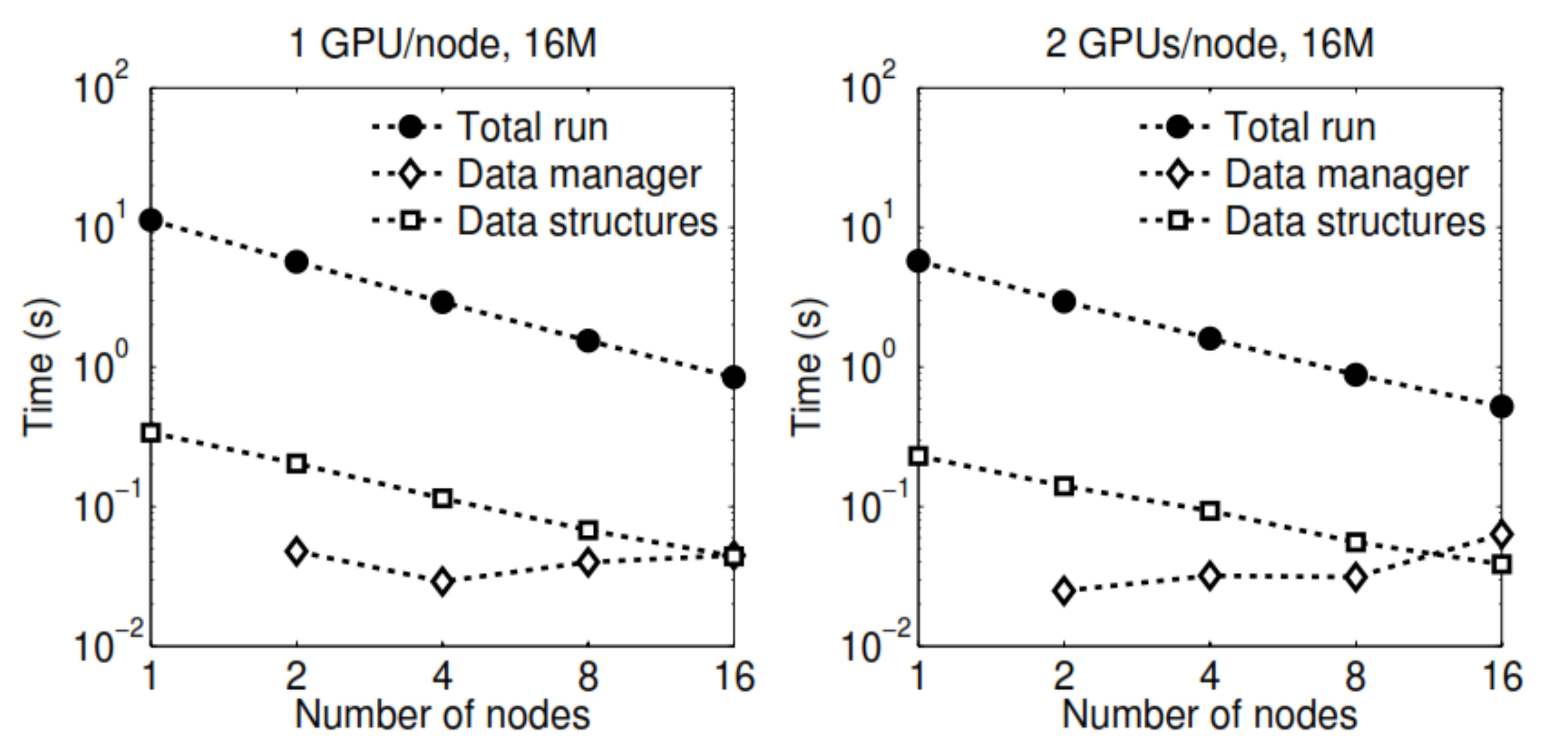}
\end{center}
\caption{The data manager and data structure processing time against
the total run time. The problem sizes are fixed to be 16M running on 1
to 16 nodes. Each node uses 1 (left) or 2 (right) GPUs. The time is measured for potential only computations.}
\label{dataManager}
\end{figure}


We used a small cluster (``Chimera'') at the University of
Maryland for tests, which has 32 nodes interconnected
via Infiniband. Each node was composed of a dual socket
quad-core Intel Xeon X5560 2.8 GHz CPUs, 24 GB\ of RAM\, and two Tesla
C1060 GPUs. We define {\it concurrent region} here as the period when
the GPU(s) computes local summation and  the CPU cores compute translation
simultaneously. In all tests we used the FMM for the Laplace equation
in 3D, $\Phi (\mathbf{y},\mathbf{x})=1/|\mathbf{y}-\mathbf{x}|$ (see
\cite{Gumerov05:JCP}).

First, the weak scalability of our algorithm was tested by
fixing the number of particles per node to $N/P=2^{23}$ and varying
the number of nodes. In Fig.~\ref{overheadComp}, we show
our overhead vs. concurrent region time against the baseline algorithm
performance. For perfect parallelization/scalability, the run
time in this case should be constant. In practice, we observed an
oscillating pattern with slight growth of the average time. In
\cite{Qi11:SC11}, two factors were explained which affect the perfect
scaling: reduction of the parallelization efficiency of the CPU part
of the algorithm and the data transfer overheads, which also applies
to our results. With the box type information, we could fully
distribute all the translations among nodes and avoid the 
unnecessary duplication of the data structure, which would become
significant at large sizes. Since our import/export data of each node
only relates to the boundary surfaces, we improve the deficiency of
their simplified algorithm that also shows up in the data transfer
overheads, which increases with $l_{\max }$.  

In Fig.~\ref{overheadComp}, the full FMM algorithm shows almost the
same CPU/GPU concurrent region time for the cases with similar particle density (the average
number of particles in a spatial box at $l_{max}$).Moreover, the
overheads of this algorithm only slightly increases in contrast to the
big jump seen the baseline algorithm when $l_{max}$ changes. Even though
the number of particles on each node remains the same, the problem
size increases hence results in the deeper octree and more spatial
boxes to handle, which also contributes to such overhead increase
(besides communication cost). Nevertheless, it could improve the
overall algorithm performance and its weak scalability by using these new data
structures.

Second, we also performed the strong scalability test, in which
$N$ is fixed and $P$ is changing (Fig.~\ref{strongScale}). The tests
were performed for $N=2^{23},2^{24}$ and $ P=1,2,4,8,16,32$ with one and
two GPUs per node. Even though, our algorithm demonstrates superior
scalability compared with the baseline algorithm, we still observe the
slight deviations from the perfect scaling for the 8M case. For 16M
case, the total run time of both 1 and 2 GPU shows the well scaling
because the GPU work was a limiting factor of CPU/GPU concurrent region
(the dominant cost). This is consistent with the fact that the sparse
MVP alone is well scalable.  For 8M case, in the case of two GPUs, the CPU work was a limiting factor for the parallel
region. However, we can see approximate correspondence of the times
obtained for two GPUs/node to the ones with one GPU/node,
i.e. doubling of the number of nodes with one GPU or increasing the
number of GPUs results in approximately the same timing. This shows a
reasonably good balance between the CPU and GPU work in the case of 2
GPUs per node, which implies this is more or less the optimal configuration for a given problem size.

Finally, we validated the acceptable cost of the communication scheme and the
computation of box type. There is a {\em data manager} in
our multiple node FMM algorithm, which is used to manage the box
import/export and communications. So we can compare this data manager processing time
(including M-data exchange) and the overall data structure
construction time with the total running time in
Fig.~\ref{dataManager}. Given the problem size and truncation number
fixed, our communication increases as the number of nodes (roughly
$P^{1/3}$ in Eq.~\ref{eqn3}). In our strong scalability tests, such
time is in the order of 0.01 seconds while the wall clock time is in
the order of 1 or 0.1 seconds (contribute $1\%\sim15\%$ of overall
time), even though GPUs are not fully occupied in some cases. This implies
such cost can be neglected in larger problems, in which the kernel
evaluations keep all GPUs fully loaded. Our implementation incorporate
the box type computation with other data structures, such as octree
and translation neighbors, hence it makes more sense to report the
total data structure cost. From Fig.~\ref{dataManager} we observe that
our data structure time decrease similarly as the wall clock time (as
$1/P$) and shows good strong scalability.

However, it could be problematic if each node is only assigned a small number of boxes, which
would occur given a large number of nodes. Eventually the subdivision of the domain would result
in the number of boxes in the boundary region of each sub-domain is
more or less the same as that of domain itself. In this
case, the number of boxes to exchange is almost the same as the total global spatial boxes. Note that although
the data manager processes the box data searching and consolidating, its main
cost comes from the communication but not those processing. Hence, all the traffic (each box has $2p^2$
coefficients) that must go through the master node will become the bottleneck of
the entire system. However, this communication traffic issue is intrinsic to the splitting
of the global octree. One possible mitigation might be implementing
a many-to-many communication model. In fact, in our current
implementation, each node is capable of computing the sending or requesting
address (node IDs) of each export or import box and this further
improvement by investigating communication cost is left for future
work.

\section{Conclusion}

For the single-node FMM, we are able to device a new algorithm, which also has the advantage that it
achieves the FMM\ data structure in $O(N)\;$time, bringing the overall
complexity of the FMM\ to this level for a given accuracy. Comparison of the GPU and CPU
times for the same algorithm show accelerations in the range 20-100 times. This
shows the feasibility of the use of GPUs for data structure
construction, which satisfyingly reduce the data-structure step to a
small part of the FMM\ overall computation time. The multiple node data structures developed here can handle non-uniform
distributions and achieve workload balance. We developed
parallel algorithms to determine the import and export boxes in which
the granularity is spatial boxes. Their parallel GPU implementations
are shown to have very small overhead and good scalability.


\subsection*{Acknowledgements}
Work partially supported by AFOSR under MURI Grant
W911NF0410176 (PI Dr.~J.~G. Leishman, monitor Dr.~D. Smith); in
addition NG was partially supported by Grant G34.31.0040 (PI
Dr.~I.~Akhatov) of the Russian Ministry of Education \& Science. We
acknowledge NSF award 0403313 and NVIDIA for the Chimera cluster at
the CUDA Center of Excellence at UMIACS. Work also partially supported by Fantalgo, LLC;

\bibliographystyle{elsarticle-num}
\bibliography{FMM_Bib}

\end{document}